\documentclass[lettersize,journal]{IEEEtran}
\usepackage{amsmath,amsfonts}
\usepackage{amssymb}
\usepackage{algorithmic}
\usepackage{algorithm}
\usepackage{array}
\usepackage{gensymb}
\usepackage{textcomp}
\usepackage{stfloats}
\usepackage{url}
\usepackage{verbatim}
\usepackage{graphicx}
\usepackage{subcaption}
\usepackage{cite}

\usepackage[absolute,showboxes]{textpos}

\TPMargin{5pt}

\newcommand{\copyrightstatement}{
    \begin{textblock}{0.84}(0.06,0.01)    
         \noindent
         \footnotesize
         \copyright  2026 IEEE.  Personal use of this material is permitted.  Permission from IEEE must be obtained for all other uses, in any current or future media, including reprinting/republishing this material for advertising or promotional purposes, creating new collective works, for resale or redistribution to servers or lists, or reuse of any copyrighted component of this work in other works. The peer-reviewed paper is available at \url{https://doi.org/10.1109/LWC.2026.3704614}.
    \end{textblock}
}

\DeclareMathOperator{\vect}{\mathrm{vec}}

\begin{document}
\copyrightstatement
\bstctlcite{IEEEexample:BSTcontrol}
\title{Data-Aided Channel and Doppler Estimation for mMIMO LEO SatComs with Uncompensated Doppler}

\author{Abdollah~Masoud~Darya,~\IEEEmembership{Graduate~Student~Member,~IEEE,}
        and~Saeed~Abdallah,~\IEEEmembership{Senior~Member,~IEEE}
\thanks{Abdollah Masoud Darya is with SSAH and the Department of Electrical Engineering, University of Sharjah, Sharjah, UAE (email: abdollah.masoud@ieee.org). Saeed Abdallah (corresponding author) is with the Department of Electrical Engineering, University of Sharjah, Sharjah, UAE (email: sabdallah@sharjah.ac.ae).}
}


\maketitle

\begin{abstract}
This paper presents a framework for estimating and tracking massive multiple-input multiple-output (mMIMO) low-Earth-orbit (LEO) satellite channels under uncompensated Doppler. The approach begins with a pilot-based minimum mean square error (MMSE) estimate, followed by Doppler estimation and data-aided channel estimation using either a decision-directed MMSE (DD-MMSE) or an expectation-maximization (EM)-based estimator. The proposed framework achieves improved channel and Doppler estimation accuracy compared to existing methods. Results demonstrate that the DD-MMSE variant offers lower complexity, while the EM variant provides higher estimation accuracy.
\end{abstract}


\section{Introduction}\label{intro}
\IEEEPARstart{A}{new} generation of user-centric satellite communications is taking shape. Namely, direct-to-cellular (D2C) connectivity that is enabled by large constellations of low-Earth-orbit (LEO) satellites with massive phased array antennas \cite{andrews20246g}. These massive multiple-input multiple-output (mMIMO)  antenna arrays contain thousands of elements that leverage several key technologies, such as adaptive beamforming \cite{Zhao2025Iterative}, to provide global connectivity, which is key in fulfilling the need for ubiquitous connectivity \cite{andrews20246g}.\par
A key issue in satellite communications is the estimation of the time-varying channel state information (CSI). Most notably, due to the high speed and altitude of LEO satellites, uncompensated Doppler effects and long propagation delays induce a phenomenon known as channel aging \cite{darya2024semi}. Channel aging causes pilot-based channel estimates to become obsolete, necessitating frequent transmissions of pilot sequences, which adds more overhead and limits the system's data rate.\par
To address this issue, previous work considered either tracking the time-varying CSI or channel parameters such as the Doppler shift, using a Kalman filter based approach \cite{yue2022block,lin2024exploiting}. However, channel/parameter tracking methods either assume prior knowledge of the user terminal's (UT) accurate position, the satellite's orbit characteristics, or perfect channel knowledge at the pilot phase, combined with a slow-varying channel (due to perfect/near-perfect Doppler compensation). This assumption does not hold in scenarios where Doppler pre-compensation is impractical, leading to rapidly time-varying channels. This can occur when positioning solutions provided by global navigation satellite systems (GNSS) are degraded due to GNSS jamming or spoofing. Furthermore, Doppler pre-compensation is infeasible when UTs do not possess a GNSS receiver due to cost or power constraints.\par
In \cite{Darya2026EM}, the authors proposed data-aided channel estimators for the case where the UT's position and velocity are not accurately known. However, that approach assumed that the satellite's real-time position is perfectly known. Furthermore, they used pilot-based CSI estimates to initialize the expectation-maximization (EM) algorithm. However, when there is uncertainty in the satellite's position, its Doppler shift cannot be known and is therefore not compensated. This induces significant time variations into the CSI, making pilot-based CSI estimates obsolete, and preventing them from being used to initialize EM.\par
This work proposes a framework to estimate and track CSI in scenarios where neither the satellite nor the receiver has accurate spatial information, resulting in uncompensated Doppler shifts. The framework uses an initial estimate from a pilot-based minimum mean square error (PB-MMSE) estimator, followed by estimation and compensation of the Doppler shift, as well as prediction of the channel for future time slots. This prediction is used to initialize either a novel low-complexity decision-directed MMSE (DD-MMSE) estimator or a modified EM estimator that incorporates pilot symbols and soft decisions. The main contribution of this work lies not in the individual estimators, but in their integration into a unified framework designed for LEO mMIMO systems with uncompensated Doppler. In particular, this work proposes combining Doppler estimation, channel prediction, data-aided estimation, and regularization to address severe channel aging without relying on prior position information or Doppler pre-compensation. The performance of the proposed framework in terms of Doppler estimation is compared to the estimation of signal parameters via rotational invariance techniques (ESPRIT) method from \cite{lin2024exploiting} and the correlation-based (CB) method from \cite{vincent2023An}, while the channel estimation performance is compared to a Kalman-based channel tracking approach from \cite{yue2022block}.\par
The remainder of this paper is organized as follows. Section~\ref{ChannelModel} describes the mMIMO LEO satellite system and channel model. The proposed channel estimation and tracking framework is introduced in Section~\ref{method}, while its performance is evaluated and discussed in Section~\ref{RnD}. Concluding remarks are provided in Section~\ref{conc}.\par

\section{System Model}\label{ChannelModel}
We study a mMIMO LEO satellite communications system that concurrently serves $K$ UTs, each equipped with a single antenna. The satellite is equipped with a uniform planar array (UPA) with a total of $M=M_xM_y$ elements, arranged along the $x$ and $y$ dimensions. Multi-beam transmission with full frequency reuse is assumed. The system operates under the frequency division duplexing mode and uses orthogonal frequency-division multiplexing (OFDM) \cite{you2020massive}.\par
For time $t$ and frequency $f$, the uplink channel between the satellite and UT $k$ is modeled as a combination of the line of sight (LoS) and non-line of sight (NLoS) components \cite{yue2022block}, and can be expressed as
\begin{equation}\label{eq1}
{h}_k(t,f)=\sqrt{\beta_k}\exp\left\{j2\pi t\nu_k^{\text{SAT}}\right\}\left(h_k^{\text{LoS}}(t,f)+h_k^{\text{NLoS}}(t,f)\right).
\end{equation}
The large-scale fading coefficient ${\beta_k}$ per UT $k$ is given by ${\beta_k}=\left(\frac{\mathfrak{c}}{4\pi d_k f_c}\right)^2$, where $\mathfrak{c}$ is the speed of light, $f_c$ is the carrier frequency, and $d_k$ represents the satellite--UT separation \cite{bjornson2024introduction}. The term $\nu_k^{\text{SAT}}$ denotes the Doppler shift induced by the satellite motion.\par 
The corresponding LoS and NLoS channel components are written as \cite{zhang2022deep}
{\scriptsize
\begin{equation}
\begin{split}
h_k^{\text{LoS}}(t,f)=&\sqrt{\frac{\kappa_k}{\kappa_k+1}}\exp\left\{j2\pi(t\nu_k^{\text{UT-LoS}}-f\tau_k^{\text{LoS}})\right\},\\
h_k^{\text{NLoS}}(t,f)=&\sqrt{\frac{1}{\kappa_k+1}}\sum_{p=1}^{P_k}g_{k,p}\exp\left\{j2\pi(t\nu_{k,p}^{\text{UT-NLoS}}-f\tau_{k,p}^{\text{NLoS}})\right\},
\end{split}
\end{equation}}
where, per user $k$ and path $p$, $g_{k,p}\sim\mathcal{C}\mathcal{N}(0,\gamma_{k,p})$ denotes the Rayleigh fading gain, where $\sum_p\gamma_{k,p}=1$ and $\{\gamma_{k,p} \mid \forall p\}$ is the power delay profile of UT $k$ \cite{li2023channel}, and $\kappa_k$ is the Rician factor. Furthermore, $\tau_k^{\text{LoS}}$ and $\tau_{k,p}^{\text{NLoS}}$ are the LoS and NLoS propagation delays, $\nu_k^{\text{UT-LoS}}$ and $\nu_{k,p}^{\text{UT-NLoS}}$ are the LoS and NLoS user Doppler shifts, and $P_k$ denotes the number of NLoS propagation paths, respectively. Note that $\tau_{k,p}^{\text{NLoS}}=\tau_k^{\text{LoS}}+\tau_{k,p}^{\text{MP}}$, where $\tau_{k,p}^{\text{MP}}$ is the multipath time delay generated according to an exponential distribution that is truncated at $\tau_\text{max}^{\text{MP}}$, with the mean set as the root-mean-square delay spread $\tau_\text{rms}$ \cite{3gpp}.\par
The demodulated uplink signal received on subcarrier $c$ during OFDM symbol $s$ can be written as \cite{you2020massive} 
\begin{equation}
\label{eq6}
\boldsymbol{y}_{s,c}= \sum_{k=1}^{K}{h}_{k,s,c}\boldsymbol{a}_{k}x_{k,s,c}+\boldsymbol{z}_{s,c} \in\mathbb{C}^{M\times 1},
\end{equation}
where the symbol transmitted by UT $k$ is denoted by $x_{k,s,c} \in \mathbb{C}$, ${h}_{k,s,c}=h_{k}\left(sT_{sd},cf_s\right)$, and the additive white Gaussian noise is denoted by $\boldsymbol{z}_{s,c} \sim\mathcal{C}\mathcal{N}\left(\boldsymbol{0},\sigma_n^2\boldsymbol{I}\right)$. Moreover, the UPA response vector is denoted by \cite{you2020massive}
\begin{equation}
\boldsymbol{a}_{k} = \boldsymbol{v}_x\left(\sin\theta^y_k\cos\theta^x_k\right) \otimes \boldsymbol{v}_y\left(\cos\theta^y_k\right) \in \mathbb{C}^{M\times 1}~,
\end{equation}
where $\theta^y_k,\,\theta^x_k \sim \mathcal{U}(\pi/2-\vartheta_{\max},\,\pi/2+\vartheta_{\max})$ are the angles associated with the $x$- and $y$-axes of the propagation paths for UT $k$, respectively, $\mathcal{U}(a,b)$ denotes the continuous uniform distribution over the interval $[a,b]$, $\vartheta_{\max}$ is the maximum UT nadir angle, and $\otimes$ represents the Kronecker product \cite{You2024Integrated}.

For $d\in\left\{x,y\right\}$, the array vector $\boldsymbol{v}_d\left(\mathcal{D}\right) \in \mathbb{C}^{M_d\times 1}$ is denoted by
\begin{equation}
\boldsymbol{v}_d\left(\mathcal{D}\right)\!=\!\frac{\left[1, \exp\left\{-j\pi\mathcal{D}\right\}, \cdots, \exp\left\{-j\pi\left(M_d-1\right)\mathcal{D}\right\}\right]^T}{\sqrt{M_d}}.
\end{equation}
The total duration of an OFDM symbol, denoted by $T_{sd}$, is represented by the sum of the cyclic prefix duration and the product of the subcarrier count $N_{sc}$ and the sampling interval $T_s$, expressed as $T_{sd}=N_{sc}T_s+T_{cp}$. The cyclic prefix duration is represented by the product of the number of cyclic prefix samples $N_{cp}$ and $T_s$, i.e., $T_{cp}=N_{cp}T_s$. Furthermore, the subcarrier index $c$ is denoted by $c=f-0.5(N_{sc}+1)$ for $f=1,\dots,N_{sc}$, while the subcarrier spacing is represented by $f_s$ \cite{You2024Integrated}.\par
In accordance with the motivations presented in Section \ref{intro}, the system model assumes that neither the satellite nor the UTs possess precise, instantaneous knowledge of their spatial coordinates. However, due to the multi-beam nature of the satellite, it can locate the users to within a beam width \cite{pat1}, which enables the satellite to calculate the vector $\boldsymbol{a}_{k}$. Furthermore, due to the unreliable satellite and user position data, the Doppler components $\{\nu_k^{\text{SAT}},\nu_{k,p}^{\text{UT-LoS}},\nu_{k,p}^{\text{UT-NLoS}}\}$ will remain uncompensated. These uncompensated effects will accelerate channel aging and complicate the channel estimation process. Therefore, this work proposes a framework that aims to reliably estimate the channel in this scenario.\par

\section{Methodology}\label{method}

This section begins by introducing the PB-MMSE approach in Subsection \ref{PMMSE}, followed by the data-aided estimators in Subsections \ref{DDMMSE} and \ref{EM}. Additionally, the complexity of the proposed estimators will be discussed in Subsection \ref{complex}.\par
\subsection{Pilot-Based MMSE Estimator}\label{PMMSE}
Let UTs $k=[1,\ldots,K]$ transmit a sequence of $s=[1,\dots,S]$ Zadoff-Chu pilot symbols followed by $\acute{s}=[S+1,\ldots,\acute{S}]$ data symbols, omitting $c$ for brevity.
The PB-MMSE estimator can then be represented as \cite{bjornson2024introduction}
\begin{equation}\label{MMSE}
    \hat{h}_{k,s}^\mathrm{MMSE}=\frac{\sigma^2_hx_{k,s}^*}{\sigma^2_h\vert x_{k,s}\vert^2+\sigma^2_n}\tilde{y}_{k,s},
\end{equation}
where $x_{k,s}$ is the transmitted Zadoff-Chu pilot symbol $s$ by UT $k$, and the slow-varying parameters $\sigma^2_h=\mathbb{E}\left[\beta_k\right]$ and $\sigma^2_n$ are perfectly known by the satellite and UTs \cite{li2023channel}. Furthermore, $\tilde{y}_{k,s}$ is a single element of $\tilde{\boldsymbol{y}}_{s,c}=[\boldsymbol{a}_{1},\ldots,\boldsymbol{a}_{K}]^{\dagger}\cdot\boldsymbol{y}_{s,c}\in\mathbb{C}^{K\times 1}$, i.e., for user $k$, where ${\dagger}$ represents the pseudo-inverse operation.\par
Assuming a LoS-dominated Rician channel ($\kappa_k\gg1$) with negligible Doppler and delay spread \cite{you2020massive}, and a time-invariant complex gain $\alpha_k$ over the observation window. Then, to facilitate Doppler estimation, a low-dimensional representation of \eqref{eq1} is shown as
\begin{equation}\label{general}
\mathfrak{h}_{k,s}\approx \alpha_k\exp\left\{j2\pi sT_\mathrm{sd} \tilde{\nu}_{k}\right\},
\end{equation}
in the time domain\footnote{In the frequency domain $\mathfrak{h}_{k,c}\approx \alpha_k\cdot\exp\left\{j2\pi cf_s \tilde{\tau}_k\right\}$.}. The complex gain is represented as $\alpha_k=\vert\alpha_k\vert\exp\{j\phi_k\}$ with initial phase offset $\phi_k$, and the effective Doppler component is represented as $\tilde{\nu}_k\approx \nu_k^{\text{SAT}}+\delta_k$. Note that $\tilde{\nu}_k$ is assumed constant within a single uplink frame due to typical LEO Doppler variation rates on the order of hundreds of Hz/s \cite{3gpp}. The residual $\delta_k$ accounts for user-induced Doppler and satisfies $\vert \delta_k\vert\ll\vert\nu_k^{\text{SAT}}\vert$, i.e., $\tilde{\nu}_k$ is dominated by the satellite motion. It is important to note that \eqref{general} does not imply a strict single-path channel. Rather, it represents a low-dimensional approximation of the multipath channel in \eqref{eq1}, which is justified under the LoS-dominated conditions of the LEO channel and the dominance of a common Doppler component induced by satellite motion. Under these conditions, the combined effect of the multipath components can be captured through $\alpha_k$ and $\tilde{\nu}_k$. Residual multipath effects are mitigated through the proposed data-aided estimation frameworks.\par
The phase of the channel estimate from \eqref{MMSE} can be related to $\tilde{\nu}_k$ through \cite{huang2022phase}
\begin{equation}
    \angle\hat{h}_{k,s}\approx 2\pi sT_\mathrm{sd}\hat{\tilde{\nu}}_{k}+\hat{\phi}_{k}.
\end{equation}
Furthermore, $\hat{\tilde{\nu}}_{k}$ is the estimated Doppler, which can then be deduced through the LS formulation
\begin{equation}
    \min_{\tilde{\nu}_k,\phi_k}\sum_{s=1}^{S}\vert\angle\hat{h}_{k,s}-(2\pi sT_\mathrm{sd}{\tilde{\nu}}_{k}+\phi_{k})\vert^2,
\end{equation}
which is equivalent to
\begin{equation}\label{EqDopp}
[2\pi\hat{\tilde{\nu}}_{k},\hat{\phi}_{k}]^T=\angle\hat{\boldsymbol{h}}_{k}\left[[T_\mathrm{sd},\ldots,ST_\mathrm{sd}]^T,\boldsymbol{1}_{S\times 1}\right]^\dagger,
\end{equation}
where $\hat{\boldsymbol{h}}_{k}=[\hat{h}_{k,1},\ldots,\hat{h}_{k,S}]^T$. Then, by compensating the estimated Doppler $\hat{\tilde{\nu}}_{k}$ from the estimated channel $\hat{h}_{k,s}$, $\hat{\alpha}_k$ can be estimated as
\begin{equation}\label{EqAlph}
 \hat{\alpha}_k=\frac{1}{S}\sum_{s=1}^S\hat{h}_{k,s}\exp\{-j2\pi sT_\mathrm{sd}\hat{\tilde{\nu}}_k\}.   
\end{equation}
The estimation of the $\hat{h}_{k,s}$ and consequently $\hat{\tilde{\nu}}_{k}$ and $\hat{\alpha}_k$ parameters enables the prediction of the future channel
\begin{equation}\label{EqProj}
    \hat{\bar{\mathfrak{h}}}_{k,\acute{s}}^\mathrm{MMSE}=\hat{\alpha}_k\exp\{j2\pi \acute{s}T_\mathrm{sd} \hat{\tilde{\nu}}_k\}.
\end{equation}
This is crucial for the next step, where an initial channel estimate is required for the equalization and detection of data symbols for the DD estimator. It is also used as an initial estimate for the EM estimator, where pilot-based estimates would not suffice due to rapid channel aging. In these scenarios, the predicted channel estimate $\hat{\bar{\mathfrak{h}}}_{k,\acute{s}}^\mathrm{MMSE}$ can be used.\par

\subsection{Decision Directed MMSE Estimator}\label{DDMMSE}
First, an MMSE equalizer is used to estimate the data symbols, using the predicted estimate from \eqref{EqProj}. The equalizer is represented by
\begin{equation}
    \hat{x}_{k,\acute{s}}=\frac{(\hat{\bar{\mathfrak{h}}}_{k,\acute{s}}^\mathrm{MMSE})^{*}}{\vert \hat{\bar{\mathfrak{h}}}_{k,\acute{s}}^\mathrm{MMSE}\vert^2+\sigma^2_n}\tilde{y}_{k,\acute{s}}.
\end{equation}
Then, for detection, a minimum distance detector is used \cite{zakharov2009optimal}, which results in the detected data symbol $\hat{\bar{x}}_{k,\acute{s}}$. The detected data symbols, combined with the known pilot symbols ${x_{k,s}}$, can then be used to obtain an updated channel estimate. This method is known as the DD-MMSE estimator. While a least-squares DD estimator was proposed in \cite{darya2024semi}, this work proposes DD-MMSE, which leverages channel and noise statistics for better resilience against noise in low SNR scenarios. The proposed DD-MMSE estimator can be formulated as
\begin{equation}
\hat{h}_{k,\breve{s}}^\mathrm{DD}=\frac{\sigma^2_h\hat{\bar{x}}_{k,\breve{s}}^*}{\sigma^2_h\vert \hat{\bar{x}}_{k,\breve{s}}\vert^2+\sigma^2_n}\tilde{y}_{k,\breve{s}},
\end{equation}
where $\breve{s} \in \{1,\ldots,\acute{S}\}$. Then, following the procedure presented in \eqref{EqDopp}--\eqref{EqProj}, $\hat{\tilde{\nu}}_k$ can be estimated, and the accuracy of the channel projection obtained in \eqref{EqProj} can be improved. Furthermore, to achieve better performance, the formulation in \eqref{EqAlph} can be replaced with
\begin{equation}\label{EqAlph2}
\hat{\boldsymbol{\alpha}}_k=\boldsymbol{\psi}\boldsymbol{\psi}^T\left(\hat{\boldsymbol{h}}_{k}^\mathrm{DD}\odot\exp\{-j2\pi\hat{\tilde{\nu}}_k[T_\mathrm{sd},\ldots,\acute{S}T_\mathrm{sd}]\}\right).   
\end{equation}
where $\odot$ is the Hadamard product, $\boldsymbol{\psi}=[\psi(1),\ldots,\psi(\acute{S})]^T\in\mathbb{R}^{\acute{S}\times 2}$, and 
\begin{equation}
\psi(\breve{s})=\left[\frac{1}{\sqrt{\acute{S}}}, \frac{1-\frac{2(\breve{s}-1)}{\acute{S}-1}}{\sqrt{\frac{\acute{S}(\acute{S} + 1)}{3(\acute{S} - 1)}}}\right]^T.
\end{equation}
The first and second columns in $\boldsymbol{\psi}$ represent the first and second Legendre polynomial basis functions \cite{senol2012nondata}. The product $\boldsymbol{\psi}\boldsymbol{\psi}^T$ is used in this work as a regularization metric which removes high frequency estimation errors but maintains the temporal variations in $\hat{\boldsymbol{\alpha}}_k$ due to the residual uncompensated Doppler effects\footnote{These effects are due to the generalization made in \eqref{general}.} \cite{Darya2026EM}.\par
Then the channel estimate can be reconstructed as
\begin{equation}
\hat{\bar{\boldsymbol{\mathfrak{h}}}}_{k}^\mathrm{DD}=\hat{\boldsymbol{\alpha}}_k\odot\exp\{j2\pi\hat{\tilde{\nu}}_k[T_\mathrm{sd},\ldots,\acute{S}T_\mathrm{sd}]\}.
\end{equation}
A summary of the framework is provided in Algorithm \ref{alg1}.\par

\begin{algorithm}[!t]
\caption{Proposed DD-MMSE-based Framework}
\begin{algorithmic}[1]
\renewcommand{\algorithmicrequire}{\textbf{Input:}}
\renewcommand{\algorithmicensure}{\textbf{Output:}}
{\small \REQUIRE PB-MMSE channel prediction $\hat{\bar{\mathfrak{h}}}_{k,\acute{s}}^\mathrm{MMSE}$}
{\small \ENSURE Updated DD-MMSE channel estimate $\hat{\bar{\boldsymbol{\mathfrak{h}}}}_{k}^\mathrm{DD}$}
\\{\scriptsize  \hspace*{-\algorithmicindent}%
$\forall k \in \{1,\ldots,K\},\ \forall \acute{s} \in \{S+1,\ldots,\acute{S}\},\ \forall \breve{s} \in \{1,\ldots,\acute{S}\}:\ $}
\STATE {\scriptsize $\hat{x}_{k,\acute{s}}=((\hat{\bar{\mathfrak{h}}}_{k,\acute{s}}^\mathrm{MMSE})^{*}\tilde{y}_{k,\acute{s}})/(\vert \hat{\bar{\mathfrak{h}}}_{k,\acute{s}}^\mathrm{MMSE}\vert^2+\sigma^2_n)$ \hfill\COMMENT{Equalization}}
\STATE {\scriptsize Detect $\hat{x}_{k,\acute{s}} \Rightarrow \hat{\bar{x}}_{k,\acute{s}}$ \hfill\COMMENT{Detection}}
\STATE {\scriptsize $\hat{\bar{x}}_{k,\breve{s}} =
\begin{cases}
x_{k,\breve{s}}, & \breve{s} \in \{1,\ldots,S\}, \\[4pt]
\hat{\bar{x}}_{k,\breve{s}}, & \breve{s} \in \{S+1,\ldots,\acute{S}\}.
\end{cases}$ \hfill\COMMENT{Combine Pilots and Data}}
\STATE {\scriptsize $\hat{h}_{k,\breve{s}}^\mathrm{DD}=(\sigma^2_h\hat{\bar{x}}_{k,\breve{s}}^*\tilde{y}_{k,\breve{s}})/(\sigma^2_h\vert \hat{\bar{x}}_{k,\breve{s}}\vert^2+\sigma^2_n)$ \hfill\COMMENT{CSI Estimation}}
\STATE {\scriptsize $[2\pi\hat{\tilde{\nu}}_{k},\hat{\phi}_{k}]^T=\angle\hat{\boldsymbol{h}}_{k}^\mathrm{DD}[[T_\mathrm{sd},\ldots,\acute{S}T_\mathrm{sd}]^T,\boldsymbol{1}_{\acute{S}\times 1}]^\dagger$ \hfill\COMMENT{Doppler Estimation}}
\STATE {\scriptsize $\hat{\boldsymbol{\alpha}}_k=\boldsymbol{\psi}\boldsymbol{\psi}^T(\hat{\boldsymbol{h}}_{k}^\mathrm{DD}\odot\exp\{-j2\pi\hat{\tilde{\nu}}_k[T_\mathrm{sd},\ldots,\acute{S}T_\mathrm{sd}]\})$ \hfill\COMMENT{Gain Estimation}}
\STATE {\scriptsize $\hat{\bar{\boldsymbol{\mathfrak{h}}}}_{k}^\mathrm{DD}=\hat{\boldsymbol{\alpha}}_k\odot\exp\{j2\pi\hat{\tilde{\nu}}_k[T_\mathrm{sd},\ldots,\acute{S}T_\mathrm{sd}]\}$ \hfill\COMMENT{CSI Reconstruction}}%
\end{algorithmic}
\label{alg1}
\end{algorithm}

\subsection{Data-Aided EM Estimator}\label{EM}
Let the posterior probability \cite{Darya2026EM}
\begin{equation}\label{posterior}
P(\xi_{\mathfrak{n}}|\tilde{y}_{k,\acute{s}},\hat{\bar{\mathfrak{h}}}_{k,\acute{s}}^{(\imath)})=
\dfrac{\exp\left(\frac{-1}{\sigma^2}\vert\tilde{y}_{k,\acute{s}}-\hat{\bar{\mathfrak{h}}}_{k,\acute{s}}^{(\imath)}\xi_{\mathfrak{n}}\vert^2\right)}{\sum_{\acute{\mathfrak{n}}=1}^{N_{\xi}}\exp\left(\frac{-1}{\sigma^2}\vert\tilde{y}_{k,\acute{s}}-\hat{\bar{\mathfrak{h}}}_{k,\acute{s}}^{(\imath)}\xi_{\acute{\mathfrak{n}}}\vert^2\right)},
\end{equation}
represent the probability of the transmission of the hidden symbol $\xi_{\mathfrak{n}}$ for $\mathfrak{n}=1,\dots,N_{\xi}$, given $\tilde{y}_{k,\acute{s}}$ and the channel estimate $\hat{\bar{\mathfrak{h}}}_{k,\acute{s}}^{(\imath)}$, where $N_{\xi}$ is the number of possible symbol hypotheses ($N_{\xi}=16$ for $16$-QAM). Note that the initial estimate is $ \hat{\bar{\mathfrak{h}}}_{k,\acute{s}}^{(1)}=\hat{\bar{\mathfrak{h}}}_{k,\acute{s}}^\mathrm{MMSE}$. The updated EM estimate then consists of two parts.
The pilot-based part is represented as $\hat{h}_{k,s}^{P}={\tilde{y}_{k,s}}/{x_{k,s}}$, while the data-aided part is represented as
\begin{equation}
\hat{h}_{k,\acute{s}}^{(\imath)}=
\frac{\sum_{\mathfrak{n}=1}^{N_{\xi}}P(\xi_{\mathfrak{n}}|\tilde{y}_{k,\acute{s}},\hat{\bar{\mathfrak{h}}}_{k,\acute{s}}^{(\imath)})\tilde{y}_{k,\acute{s}}\xi_{\mathfrak{n}}^*}{\sum_{\mathfrak{n}=1}^{N_{\xi}}P(\xi_{\mathfrak{n}}|\tilde{y}_{k,\acute{s}},\hat{\bar{\mathfrak{h}}}_{k,\acute{s}}^{(\imath)})\vert\xi_{\mathfrak{n}}\vert^2}.
\end{equation}
Therefore, the combined updated EM estimate is presented as
\begin{equation}
\hat{\boldsymbol{h}}_{k}^{(\imath)}=\left[\hat{h}_{k,1}^{P},\ldots,\hat{h}_{k,S}^{P},\hat{h}_{k,S+1}^{(\imath)},\ldots,\hat{h}_{k,\acute{S}}^{(\imath)}\right].
\end{equation}
Next, $\hat{\tilde{\nu}}_k^{(\imath)}$ is found using \eqref{EqDopp} and used in
\begin{equation}
\begin{split}
\hat{\boldsymbol{\alpha}}_k^{(\imath)}=\boldsymbol{\psi}\boldsymbol{\psi}^T\!\left(\hat{\boldsymbol{h}}_{k}^{(\imath)}\odot\exp\{-j2\pi\hat{\tilde{\nu}}_k^{(\imath)}[T_\mathrm{sd},\ldots,\acute{S}T_\mathrm{sd}]\}\right).
\end{split}
\end{equation}
Finally, the reconstructed channel estimate is defined as
\begin{equation}
\hat{\bar{\boldsymbol{\mathfrak{h}}}}_{k}^{(\imath+1)}=\hat{\boldsymbol{\alpha}}_k^{(\imath)}\odot\exp\{j2\pi\hat{\tilde{\nu}}_k^{(\imath)}[T_\mathrm{sd},\ldots,\acute{S}T_\mathrm{sd}]\},
\end{equation}
then EM estimation is performed iteratively over $\imath=1,\ldots,N_\text{EM}$. The framework is summarized in Algorithm \ref{alg2}.\par

\begin{algorithm}[!t]
\caption{Proposed EM-based Framework}
{\setlength{\baselineskip}{1.25\baselineskip}
\begin{algorithmic}[1]
\renewcommand{\algorithmicrequire}{\textbf{Input:}}
\renewcommand{\algorithmicensure}{\textbf{Output:}}
{\small \REQUIRE PB-MMSE channel prediction $\hat{\bar{\mathfrak{h}}}_{k,\acute{s}}^\mathrm{MMSE}$}
\ENSURE Updated EM channel estimate $\hat{\bar{\boldsymbol{\mathfrak{h}}}}_{k}^\mathrm{EM}$
\\{\scriptsize  \hspace*{-\algorithmicindent}%
$\forall k \in \{1,\ldots,K\},\ \forall s \in \{1,\ldots,S\},\ \forall \acute{s} \in \{S+1,\ldots,\acute{S}\}:\ $}
\\Initialize $\hat{\bar{\mathfrak{h}}}_{k,\acute{s}}^{(1)}=\hat{\bar{\mathfrak{h}}}_{k,\acute{s}}^\mathrm{MMSE}$
\FOR {$\imath = 1$ to $N_\text{EM}$}
\STATE {\scriptsize $\hat{h}_{k,s}^{P}={\tilde{y}_{k,s}}/{x_{k,s}}$ \hfill\COMMENT{Pilot-based CSI Estimation}}
\STATE {\scriptsize $P(\xi_{\mathfrak{n}}|\tilde{y}_{k,\acute{s}},\hat{\bar{\mathfrak{h}}}_{k,\acute{s}}^{(\imath)})=
\dfrac{\exp\left(\frac{-1}{\sigma^2}\vert\tilde{y}_{k,\acute{s}}-\hat{\bar{\mathfrak{h}}}_{k,\acute{s}}^{(\imath)}\xi_{\mathfrak{n}}\vert^2\right)}{\sum_{\acute{\mathfrak{n}}=1}^{N_{\xi}}\exp\left(\frac{-1}{\sigma^2}\vert\tilde{y}_{k,\acute{s}}-\hat{\bar{\mathfrak{h}}}_{k,\acute{s}}^{(\imath)}\xi_{\acute{\mathfrak{n}}}\vert^2\right)}$ \hfill\COMMENT{Expectation}}
\STATE {\scriptsize $\hat{h}_{k,\acute{s}}^{(\imath)}=
\dfrac{\sum_{\mathfrak{n}=1}^{N_{\xi}}P(\xi_{\mathfrak{n}}|\tilde{y}_{k,\acute{s}},\hat{\bar{\mathfrak{h}}}_{k,\acute{s}}^{(\imath)})\tilde{y}_{k,\acute{s}}\xi_{\mathfrak{n}}^*}{\sum_{\mathfrak{n}=1}^{N_{\xi}}P(\xi_{\mathfrak{n}}|\tilde{y}_{k,\acute{s}},\hat{\bar{\mathfrak{h}}}_{k,\acute{s}}^{(\imath)})\vert\xi_{\mathfrak{n}}\vert^2}$ \hfill\COMMENT{Maximization}}
\STATE {\scriptsize $\hat{\boldsymbol{h}}_{k}^{(\imath)}=[\hat{h}_{k,1}^{P},\ldots,\hat{h}_{k,S}^{P},\hat{h}_{k,S+1}^{(\imath)},\ldots,\hat{h}_{k,\acute{S}}^{(\imath)}]$ \hfill\COMMENT{Combine $\hat{h}_{k,s}^{P}$ and $\hat{h}_{k,\acute{s}}^{(\imath)}$}}
\STATE {\scriptsize $[2\pi\hat{\tilde{\nu}}_{k}^{(\imath)},\hat{\phi}_{k}]^T=\angle\hat{\boldsymbol{h}}_{k}^{(\imath)}[[T_\mathrm{sd},\ldots,\acute{S}T_\mathrm{sd}]^T,\boldsymbol{1}_{\acute{S}\times 1}]^\dagger$ \hfill\COMMENT{Doppler Est.}}
\STATE {\scriptsize $\hat{\boldsymbol{\alpha}}_k^{(\imath)}=\boldsymbol{\psi}\boldsymbol{\psi}^T\!\left(\hat{\boldsymbol{h}}_{k}^{(\imath)}\odot\exp\{-j2\pi\hat{\tilde{\nu}}_k^{(\imath)}[T_\mathrm{sd},\ldots,\acute{S}T_\mathrm{sd}]\}\right)$ \hfill\COMMENT{Gain Est.}}
\STATE {\scriptsize $\hat{\bar{\boldsymbol{\mathfrak{h}}}}_{k}^{(\imath+1)}=\hat{\boldsymbol{\alpha}}_k^{(\imath)}\odot\exp\{j2\pi\hat{\tilde{\nu}}_k^{(\imath)}[T_\mathrm{sd},\ldots,\acute{S}T_\mathrm{sd}]\}$ \hfill\COMMENT{CSI Reconstruction}}
\ENDFOR
\RETURN $\hat{\bar{\boldsymbol{\mathfrak{h}}}}_{k}^\mathrm{EM}=\hat{\bar{\boldsymbol{\mathfrak{h}}}}_{k}^{(N_\text{EM})}$%
\end{algorithmic}}
\label{alg2}
\end{algorithm}

While the proposed estimators share a common structure, there exist some key differences. Specifically, the use of soft decisions by the EM estimator compared to detected data by the DD-MMSE estimator, in addition to pilot symbols. This causes the computational complexity of the EM estimator to be higher while attaining better estimation accuracy.\par

\subsection{Complexity Analysis}\label{complex}
The complexity of the vectorized data-aided estimators is presented in terms of the $\mathcal{O}$ notation, for $\forall k \in \{1,\ldots,K\}$, $\ \forall \acute{s} \in \{S+1,\ldots,\acute{S}\}$, and $ \forall \breve{s} \in \{1,\ldots,\acute{S}\}$.\par
The most complex operation is $\tilde{\boldsymbol{y}}_{\breve{s},c}=[\boldsymbol{a}_{1},\ldots,\boldsymbol{a}_{K}]^{\dagger}\cdot\boldsymbol{y}_{\breve{s},c}$ with $\mathcal{O}(K^2M)$ for the pseudo-inverse and $\mathcal{O}(K\acute{S}M)$ for the product. The pseudo-inverse operation is computed offline and updated only when the satellite or UT positions change considerably \cite{you2020massive}, and is thus excluded from run-time complexity.\par
For both frameworks, the complexity of estimating $\hat{\tilde{\nu}}_{k}$ and $\hat{\boldsymbol{\alpha}}_k^{(\imath)}$ (excluding the product $\boldsymbol{\psi}\boldsymbol{\psi}^T$ which can be performed offline) is $\mathcal{O}(2K\acute{S})$ and $\mathcal{O}(K^2\acute{S})$, respectively.\par
For the DD-MMSE framework, the complexity of the equalization and calculation of $\hat{h}_{k,\breve{s}}^\mathrm{DD}$ is $\mathcal{O}(K\acute{S})$, each. Alternatively, for the EM framework, the complexity of calculating the posterior and estimating $\hat{h}_{k,\acute{s}}^{(\imath)}$ is $\mathcal{O}(K\acute{S}N_\xi)$, each. Therefore, the run-time complexity is  $\mathcal{O}(K\acute{S}(K+M+1))$ and $\mathcal{O}(K\acute{S}(K+M+N_\xi))$, for the DD-MMSE and EM frameworks, respectively.\par
The median execution time, computed over $10^5$ Monte Carlo iterations using MATLAB R2026a on a machine with an Intel Xeon 6226R CPU, is $0.59$ ms for the DD-MMSE estimator and $1.01$ ms for a single iteration of the EM estimator.\par

\section{Results and Discussion}\label{RnD}

\begin{table}[!t]
\caption{\small{Reference channel parameters\label{table3gpp}}}
\centering
\scriptsize
\begin{tabular}{|l|c|l|c|l|c|}
\hline
\textbf{Parameter} & \textbf{Value} & \textbf{Parameter} & \textbf{Value} & \textbf{Parameter} & \textbf{Value} \\
\hline
$B$ & $15\,\mathrm{MHz}$ & $\tau_\text{rms}$ & $250\,\mathrm{ns}$ & $M_xM_y$ & $16\times16$\\
\hline
$f_c$ & $2\,\mathrm{GHz}$ & $\tau_\text{max}^{\text{MP}}$ & $1\,\mathrm{\mu s}$ & $\varphi_\text{min},\varphi_\text{max}$ & $30\degree,90\degree$\\
\hline
$N_\text{EM}$ & $10$ & $K$ & $10$ & $\mathrm{R_e}$ & $6\,371\,\mathrm{km}$\\
\hline
$\vartheta_{\max}$ & $\pi/5$ & $H$ & $600\,\mathrm{km}$ & $\kappa_k$ & $10\,\mathrm{dB}$\\
\hline
$f_s$ & $60\,\mathrm{kHz}$ & $\vert\nu_\text{max}^{\text{SAT}}\vert$ & $48\,\mathrm{kHz}$ & Modulation & $16$-QAM\\
\hline
$P_k$ & $4$ & $S$ & $15$ & $\acute{S}$ & $90$\\
\hline
\end{tabular}
\end{table}

The simulation results presented in this section are obtained using reference system parameters from 3GPP \cite{3GPP38.821,3gpp}, as summarized in Table \ref{table3gpp}. The nadir and elevation angles for UT $k$ are found through $\vartheta_k=\arccos(\sin\theta^x_k\sin\theta^y_k)$ and $\varphi_k=\arccos\bigl(\frac{\mathrm{R_e}+H}{\mathrm{R_e}}\sin\vartheta_k\bigr)$, where $\mathrm{R_e}$ and $H$ denote Earth's radius and satellite altitude, respectively. The satellite Doppler shift is calculated as $\nu_k^{\text{SAT}}=\mathfrak{m}_k\vert\nu_\text{max}^{\text{SAT}}\vert(1-{\varphi_k}/{\varphi_\text{max}})$, with $\mathfrak{m}_k\sim\mathcal{U}\{-1,1\}$ determining the Doppler shift sign. The UT Doppler shift is calculated as $\nu_{k,p}^{\text{UT}}=\mathfrak{m}_k{f_c\mathrm{v}_k}\cos(\varrho_k)/{\mathfrak{c}}$, where $\mathrm{v}_k$ is the UT speed, $\varrho_k=\varphi_k$ for $\nu_k^{\text{UT-LoS}}$, and $\varrho_k\sim\mathcal{U}(0,\pi/2)$ for $\nu_k^{\text{UT-NLoS}}$ \cite{3gpp}. Note that the maximum UT speed is $100\,\mathrm{km/h}$ and the satellite orbital speed is $7.5\,\mathrm{km/s}$.\par
The performance of the proposed channel estimation and tracking approach is evaluated in terms of two metrics. First, the Doppler estimate $\hat{\tilde{\nu}}_k$ is evaluated against an estimate found by using the perfectly known channel in \eqref{EqDopp} in terms of the Mean Absolute Error (MAE). The MAE of the Doppler estimate $\hat{\tilde{\nu}}_k$ is found using
\begin{equation}\label{MAE}
\text{MAE}=\frac{1}{K}\sum_{k=1}^{K}\left\vert\hat{\tilde{\nu}}_k-\hat{\tilde{\nu}}_k^\text{ref}\right\vert.
\end{equation}
Second, the performance of the different channel estimators is compared against a Kalman-based approach in terms of the Normalized Mean Squared Error (NMSE). The NMSE of the estimate $\hat{\bar{\boldsymbol{\mathfrak{h}}}}_{s}$ is found using
\begin{equation}\label{NMSE}
\text{NMSE}=\frac{\left\Vert\vect\left(\left[\boldsymbol{h}_{1}^\text{ref},\dots,\boldsymbol{h}_{\acute{S}}^\text{ref}\right]\right)-\vect\left(\left[\hat{\bar{\boldsymbol{\mathfrak{h}}}}_{1},\dots,\hat{\bar{\boldsymbol{\mathfrak{h}}}}_{\acute{S}}\right]\right)\right\Vert^2}{\left\Vert\vect\left(\left[\boldsymbol{h}_{1}^\text{ref},\dots,\boldsymbol{h}_{\acute{S}}^\text{ref}\right]\right)\right\Vert^2},
\end{equation}
where $\vect(\cdot)$ represents the vectorization operation.\par

\begin{figure}[!t]
\centering
\includegraphics[width=1\linewidth]{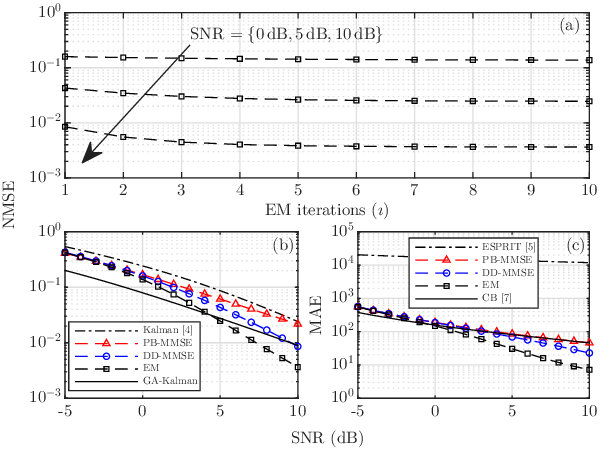}
\caption{\small{a) NMSE vs. $\imath$ of the EM estimator, b) NMSE vs. SNR of channel estimate, and c) MAE vs. SNR of Doppler estimate.}}
\label{Fig_1}
\end{figure}

Fig. \ref{Fig_1}(a) shows the performance of the EM estimator versus EM iterations $\imath$. It shows convergence within a small number of iterations. Fig. \ref{Fig_1}(b) shows the performance of the proposed estimators versus the Kalman-based approach from \cite{yue2022block}, where a symbol-by-symbol approach (block size of 1 OFDM symbol) was used for all $\acute{S}$ symbols. This is due to the rapid variation of the channel, where a larger block size (i.e., of multiple OFDM symbols) would perform very poorly.\par
Two Kalman-based trends are shown in Fig. \ref{Fig_1}(b). The Kalman approach utilizes detected data symbols (detected using $\hat{\bar{\mathfrak{h}}}_{k,\acute{s}}^\mathrm{MMSE}$), while the genie-aided Kalman (GA-Kalman) trend assumes perfectly known data symbols. The figure shows the channel prediction by the PB-MMSE estimator outperforming the Kalman-based approach. The proposed DD-MMSE and EM also outperformed the Kalman-based approach, with the EM estimator even outperforming the GA-Kalman at SNR $\geq5\,\mathrm{dB}$. This is due to EM's iterative approach and its reliance on soft decisions, compared to the hard decisions by the Kalman filter, as well as the regularization method used that eliminates high-frequency estimation errors.\par
Fig. \ref{Fig_1}(c) shows the MAE performance of the proposed methods versus the ESPRIT Doppler estimation method \cite{lin2024exploiting} and the CB method (see (13) in \cite{vincent2023An}). We note that for the ESPRIT and CB methods, the $\hat{h}_{k,s}^\mathrm{MMSE}$ channel estimate was used. With increasing SNR, the proposed estimators are shown to converge towards an MAE of $0$, with the EM leveraging its iterative nature to achieve improved accuracy. The performance of the ESPRIT method is comparatively poor. This might be because ESPRIT is more applicable in multipath channels where each path has a comparable Doppler shift \cite{Gao2025Signal}. The CB method leverages the representation in \eqref{general}, which reduces Doppler estimation to the estimation of a single frequency component. This enabled it to outperform PB-MMSE. Yet, it was outperformed by EM at SNR $>0\,\mathrm{dB}$.\par

\section{Conclusion}\label{conc}
This work introduced a framework for LEO mMIMO channel and Doppler estimation that combines pilot-based MMSE initialization, Doppler estimation, and data-aided channel estimation via DD-MMSE or EM-based estimators, targeting scenarios without prior Doppler compensation. Simulation results show that data-aided estimation, combined with regularization and Doppler estimation, substantially improves estimation accuracy relative to related work. The DD-MMSE implementation offers lower computational complexity, while the EM-based approach attains greater robustness. Future work will explore hybrid schemes that combine the proposed estimators with data-driven approaches.

\bibliographystyle{IEEEtran}
\bibliography{IEEEabrv.bib,main.bib}

\end{document}